# LEP1 – Quick

**Bodo Lampe**

Max Planck Institut für Physik
D-80805 München, P.O. Box 401212, Germany

## Abstract

The theoretical background of the electroweak precision measurements at LEP1 is reviewed. The presentation is compact but specific enough to understand all details.

# 1. The True Weinberg Angle

In the electroweak theory there are two gauge coupling constants, usually called $g_1$ and $g_2$, one for the $SU(2)_L$ gauge bosons $W^+$, $W^-$ and $W_3$ and the other one for the $U(1)_Y$ gauge boson B. $W_3$ and B mix to yield the Z boson and the photon :

$$Z = W_3 \cos\theta + B \sin\theta \qquad (1)$$

The mixing angle $\theta$ is called the Weinberg angle and can be related to the couplings $g_1$ and $g_2$ via

$$\cos\theta = \frac{g_2}{\sqrt{g_1^2 + g_2^1}} \qquad (2)$$

The same is true for the electromagnetic coupling constant e :

$$e = \frac{g_1 g_2}{\sqrt{g_1^2 + g_2^2}} \qquad (3)$$

The latter equation follows from the requirement that the photon coupling should be as in QED. Before the advent of the LEP1 results it was common to use the quantities e and $\sin^2\theta$ instead of $g_1$ and $g_2$ to fix the couplings of the electroweak theory. After LEP1 there is the Z mass, measured at a very high precision, to substitute $\sin^2\theta$ as a fundamental parameter of the standard model. Still, I will show in the following that $\sin^2\theta$ can serve as a guideline to understand the implications of the higher order corrections on the LEP1 results [1].

At first sight, eq. 1 is only a leading order relation. In order to maintain it at higher orders [2] [3], one should reinterpret all the quantities introduced so far as renormalized quantities. For example, the fields in eq. 1 should be the renormalized gauge fields. Eq. 1 defines in a sense the "true" Weinberg angle, with the proviso that this definition depends on the renormalization scheme chosen. The most common scheme is the on-shell scheme, in which particle masses are defined as propagator poles and the electromagnetic coupling is fixed to be $ie\gamma_\mu$ at zero momentum transfer, true to all orders. If not stated otherwise, the on-shell scheme will always be the basis of our discussion. Other schemes and correspondingly other definitions of $\sin^2\theta$ are however possible. In fact one can define $\sin^2\theta$ in an infinite number of ways. Some of the possible definitions will be presented in the following, because they will help to understand the qualitative features of the higher order corrections.

## 2. Definition of $\sin^2 \theta$ in terms of the vector boson masses

In leading order the vector boson masses are given by

$$m_W = \frac{1}{2} g_2 v \qquad\qquad m_Z = \frac{1}{2}\sqrt{g_1^2 + g_2^1} v \qquad (4)$$

where v is the vacuum expectation value of the Higgs field $\Phi$. Eq. 4 can be derived by inserting v into the kinetic term $(D_\mu \Phi)^+ (D_\mu \Phi)$ of the Higgs field Lagrangian, because this produces mass terms $\approx Z_\mu Z^\mu$ and $\approx W_\mu^+ W^{-\mu}$ for Z and W. Comparing eqs. 2 and 4 one finds that the cosine of the Weinberg angle is given by the ratio $\frac{m_W}{m_Z}$. To maintain this relation at higher orders one may define a quantity

$$s_W^2 = 1 - \frac{m_W^2}{m_Z^2} \qquad (5)$$

In this equation, the vector boson masses are the on-schell masses. A definition in terms of $\overline{MS}$ masses would be possible as well, but will not be pursued here. Clearly, in lowest order $\sin^2 \theta$ and $s_W^2$ are identical, but they get different higher order corrections. The difference is described by the "$\rho$–parameter" $\rho$ defined by

$$\sin^2 \theta = 1 - \frac{m_W^2}{\rho m_Z^2} = s_W^2 + c_W^2 \Delta \rho \qquad (6)$$

In eq. 6 I have introduced a quantity $\Delta \rho$ defined by $\rho = 1 + \Delta \rho$ which is understood to be small (of oneloop order). All the discussions presented here are restricted to oneloop order in the electroweak coupling constants, so that terms of order $(\Delta \rho)^2$ etc are neglected.

$\Delta \rho$ has a representation in terms of the renormalized W and Z selfenergies [4]:

$$\Delta \rho = \left. \frac{\Sigma_{ZZ}(p^2)}{m_Z^2} - \frac{\Sigma_{WW}(p^2)}{m_W^2} \right|_{p^2=0} \qquad (7)$$

The selfenergies can be calculated from the vector boson propagators fig. 1. I will not derive this relation here, but I think it is intuitively clear, because $s_W^2$ is defined in terms of the vector boson masses and these in turn are related to the self energies. The fact that this happens at zero momentum is because through the mixing the photon enters the game. Since one wants to fix the photon selfenergy at zero momentum, one has to do the same for all other selfenergies.

The calculation of the selfenergies leads to expressions which are dominated by the large masses of the Standard Model, the Higgs mass and the top quark mass, coming

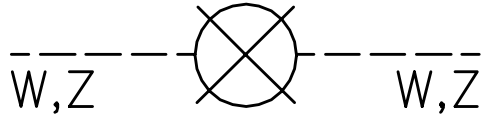

Figure 1: the crossed circle in this diagram denotes all insertions to the Z resp. W propagator

from the diagrams with a Higgs boson or a top quark in the loop. An approximate formula for $\Delta\rho$ is

$$\Delta\rho = \frac{3G_F m_t^2}{8\sqrt{2}\pi^2} - \frac{11 G_F m_W^2}{24\sqrt{2}\pi^2} \tan^2\theta \ln\left(\frac{m_H^2}{m_W^2}\right) + \ldots \qquad (8)$$

$G_F = \frac{1}{2\sqrt{2}v^2}$ is the Fermi constant as measured in muon decay and $m_H$ is the Higgs mass. Since eq. 8 is a oneloop expression it does not matter, which of the possible beyond the lo definitions of $G_F$ one chooses (see later). $G_F$ and $m_H$ are the parameters to fix the standard model Higgs Lagrangian completely. One of them ($G_F$) is known very precisely (to six digits), the other one essentially unknown. It is one of the main aims of electroweak precision studies to be able to make indirect statements about the value of $m_H$.

The dots in eq. 8 stand for (known) contributions which are smaller than the leading $m_t^2$ and $\ln(m_H)$ contributions, such as constants, vector boson and light fermion mass terms and small logarithms. [5] [6] . Eq. 8 can be viewed as the leading order of an expansion of $\Delta\rho$ in powers of $\frac{m_{W,Z}^2}{m_t^2}$ and $\frac{m_{W,Z}^2}{m_H^2}$. It can be shown that the neglected terms give a small contribution numerically. As far as I understand the literature, it is believed quite in general that the approximation $m_{W,Z} \ll m_{t,H}$ is reasonable for the higher order analysis of the LEP1 data. Still I do not recommend to trust this statement blindly, but to try to calculate the nonleading terms in $\frac{m_{W,Z}}{m_{t,H}}$ in each specific case, for safety reasons. There is another approximation which is sometimes used in electroweak higher order calculations by researchers who believe that the Higgs mass might be of the order of the W and Z mass. This would suggest the approximation $m_t \gg m_H \sim 0$. This approximation works in certain circumstances, but fails in others. There is an explicit example, in which this is not a reasonable approximation, even in case $m_H \approx m_W$. This example is a twoloop effect and will be given within the next formula (eq. 9).

Since it is a oneloop higher order effect, in eq. 8 it does not really matter which definition of $\theta$ is choosen. This is in general not true anymore if one starts to include twoloop contributions, some of which are nowadays known. In the leading $m_t$ limit $\Delta\rho$ gets an overall correction factor [7] [5] [6].

$$\Delta\rho = \frac{3G_F m_t^2}{8\sqrt{2}\pi^2}\left\{1 + \frac{G_F m_t^2}{8\sqrt{2}\pi^2}c_2 - \frac{2\alpha_s}{3\pi}(\frac{\pi^2}{3}+1)\right\} \qquad (9)$$

from twoloop contributions. The term involving $\alpha_s$ is the mixed electroweak/QCD correction. The leading twoloop electroweak correction $c_2$ has a complicated analytical form, even if $m_W$ and $m_Z$ are neglected. It becomes very simple if in addition $m_H \ll m_t$ is assumed, namely $c_2 = 19 - 2\pi^2$. However, except for very small Higgs masses, the full result deviates significantly from the approximation $m_H \approx 0$.

Another problem of the twoloop result is that it is not complete. Although some slight progress has recently been made, no estimate whatsoever of terms of the form $\alpha G_F m_t^2 \ln(m_t)$ etc exists.

Let us now study $\Delta\rho$ numerically. One finds

$$\Delta\rho = \begin{cases} 0.0100 \pm 0.002 & m_H = 60\text{GeV} \\ 0.0059 \pm 0.002 & m_H = 1000\text{GeV} \end{cases} \qquad (10)$$

This value has been obtained using $m_t = 175 \pm 20$ GeV from the Fermilab top quark analysis and a variation of $m_H$ from 60 to 1000 GeV. For an electroweak correction, this is a rather large effect coming mainly from the term of order $G_F m_t^2$. Alternatively, one can try to determine $\Delta\rho$ from the combined LEP1 data. this yields

$$\Delta\rho = 0.0084 \pm 0.0041 \qquad (\Delta\rho)_{SM} = 0.0066 \pm 0.0010 \qquad (11)$$

The value $(\Delta\rho)_{SM}$ has been obtained under the assumption that the Standard Model is correct. This induces correlations which makes the error smaller.

The genuine electroweak oneloop corrections can and have been organized in such a way that among them the $\rho$–paramter contribution is the most dominant one. In that framework $\Delta\rho$ is sometimes called $\epsilon_1$. (There are also $\epsilon_2$, $\epsilon_3$ and $\epsilon_b$ to be defined later).

What do I mean by "genuine" electroweak corrections? They are the corrections which are present beyond the "improved" Born approximation (IBA). In this work IBA is defined to contain, besides the Born term, the ordinary QED corrections (including the running of $\alpha$ between 0 and $m_Z$) and oneloop QCD corrections (in case the final state particles are quarks). Sometimes in the literature IBA is defined excluding the running of $\alpha$ and/or including the leading contribution to $\Delta\rho$. I shall mostly stick to the former definition, because it nicely isolates the pure electroweak stuff.

An important qualitative property of $\epsilon_1$ is that it is a measure of the violation of "custodial" $SU(2)_R$. Custodial $SU(2)_R$ is the righthanded global symmetry, which is broken, if weak isospin partners have different masses. For example, the Dirac mass term $m_t(\bar{t}_R t_L + \bar{t}_L t_R) + m_b(\bar{b}_R b_L + \bar{b}_L b_R)$ breaks custodial $SU(2)_R$, because $m_t \neq m_b$. In fact, the term $m_t^2$ in $\Delta\rho$ eq. 8 originates from an expression

$$m_b^2 + m_t^2 - \frac{2m_b^2 m_t^2}{m_t^2 - m_b^2} \ln \frac{m_t^2}{m_b^2} = O(m_t^2 - m_b^2) \tag{12}$$

in the limit $m_b = 0$. Similarly, the term $\ln(m_H)$ arises in eq. 8, because custodial $SU(2)_R$ is broken by the difference $m_W - m_Z$. It originates from an expression

$$\frac{m_Z^2 m_H^2}{m_Z^2 - m_H^2} \ln \frac{m_H^2}{m_Z^2} - \frac{m_W^2 m_H^2}{m_W^2 - m_H^2} \ln \frac{m_H^2}{m_W^2} \tag{13}$$

in the limit $m_H \gg m_{W,Z}$. Note that this expression vanishes in the limit $m_W = m_Z \leftrightarrow s_W^2 = 0$. The Higgs boson part of the standard model Lagrangian alone is invariant under custodial $SU(2)_R$. That is the reason, why there are no terms of order $O(m_H^2)$ in $\Delta\rho$.

### 3. Definition of $\sin^2\theta$ in terms of the muon decay constant

The most precisely known quantity in weak interaction physics today is still the Fermi constant as measured in muon decays. For that reason it should be used as one of the basic parameters of the theory. In the low energy limit the Standard Model description of the weak interaction by W and Z boson exchange should agree with the Fermi theory, which describes muon decay by an effective Lagrangian of the form

$$\frac{G_F}{\sqrt{2}}(\bar{\nu}_\mu \gamma_\lambda (1 - \gamma_5)\mu)(\bar{e}\gamma^\lambda(1 - \gamma_5)\nu_e) + h.c. \tag{14}$$

From that condition, in lowest order a relation $G_F = \frac{1}{2\sqrt{2}v^2} = \frac{e^2}{4\sqrt{2}\sin^2\theta m_W^2}$ between the Standard Model parameters and the Fermi constant can be derived. A possible definition of $\sin^2\theta$ is such that this relation is maintained to all orders [8], i.e.

$$s_0^2(1 - s_0^2) = \frac{\pi\alpha}{\sqrt{2}G_F m_Z^2} = (1 - \Delta r)(1 - \frac{m_W^2}{m_Z^2})\frac{m_W^2}{m_Z^2} \tag{15}$$

This definition of $\sin^2\theta$ by $s_0^2$ is perhaps the one which will persist in the future because it makes use of quantities which are known very precisely ($\alpha$, $G_F$ and $m_Z$). $\alpha$ is known to twelve, $G_F$ to six and $m_Z$ to five digits of accuracy. I have included in eq. 15 a second equality which gives the relation between $s_0^2$ and $s_W^2 = \frac{m_W^2}{m_Z^2}$. This relation is the defining equation for the oneloop quantity $\Delta r$ [8]. Just as $\Delta\rho$, $\Delta r$ is sensitive to $m_t$ and $m_H$. Due to the appearance of $G_F$ in eq. 15 the main ingredient to the calculation of $\Delta r$ are the higher order corrections to muon decay. These corrections involve W vacuum polarization effects, loop effects at the $\mu$–$\nu_\mu$–W and at the $e$–$\nu_e$–W vertex and boxdiagrams. It can be shown that – besides the running of $\alpha$ – the dominant contribution to $\Delta r$ comes from vacuum polarization effects, i.e. from terms which can be related to $\Delta\rho$ in the leading $m_t$ limit. In fact one can write

$$\Delta r = \Delta\alpha - \frac{c_W^2}{s_W^2}\Delta\rho + (\Delta r)_{small} \qquad (16)$$

where $(\Delta r)_{small} \leq O(0.01)$ is small as compared to $\Delta\alpha = 0.0595 \pm 0.0009$ and $\frac{c_W^2}{s_W^2}\Delta\rho$. The effect $\Delta\alpha$ from the running of $\alpha$ will be discussed in detail later. In the true spirit of the IBA it would be more convenient to remove $\Delta\alpha$ from the definition of $\Delta r$ by defining

$$\hat{s}_0^2(1-\hat{s}_0^2) = \frac{\pi\alpha(m_Z^2)}{\sqrt{2}G_F m_Z^2} = (1-\Delta\hat{r})(1-\frac{m_W^2}{m_Z^2})\frac{m_W^2}{m_Z^2} \qquad (17)$$

where $\Delta\hat{r}$ is the same as $\Delta r$ (eq. 16) except that the $\Delta\alpha$ piece is removed.

The full expression for $(\Delta r)_{small}$ is very complicated and will not be given here. Instead, I want to make explicit the connection to the calculation of the muon lifetime. To oneloop order the muon lifetime is given by

$$\frac{1}{\tau_\mu} = \frac{G_F^2 m_\mu^5}{192\pi^3}(1-\frac{8m_e^2}{m_\mu^2})\{1+\frac{\alpha(m_\mu^2)}{2\pi}(\frac{25}{4}-\pi^2)\} \qquad (18)$$

where

$$\frac{G_F}{\sqrt{2}} = \frac{e^2}{8m_W^2(1-\frac{m_W^2}{m_Z^2})}(1+\frac{\Sigma_{WW}(0)}{m_W^2}-\delta_{VB}) \qquad (19)$$

accounts for the genuine electroweak corrections. $\Sigma_{WW}(0)$ is the renormalized contribution from the W selfenergy diagrams (vacuum polarization) and $\delta_{VB}$ are the (small) contributions from vertex corrections and box diagrams. The combination $\frac{\Sigma_{WW}(0)}{m_W^2} - \delta_{VB}$ can be identified as $\Delta r$.

To better account for twoloop effects, eq. 16 is sometimes rewritten as

$$1-\Delta r = (1-\Delta\alpha)(1+\frac{c_W^2}{s_W^2}\Delta\rho) - (\Delta r)_{small} \qquad (20)$$

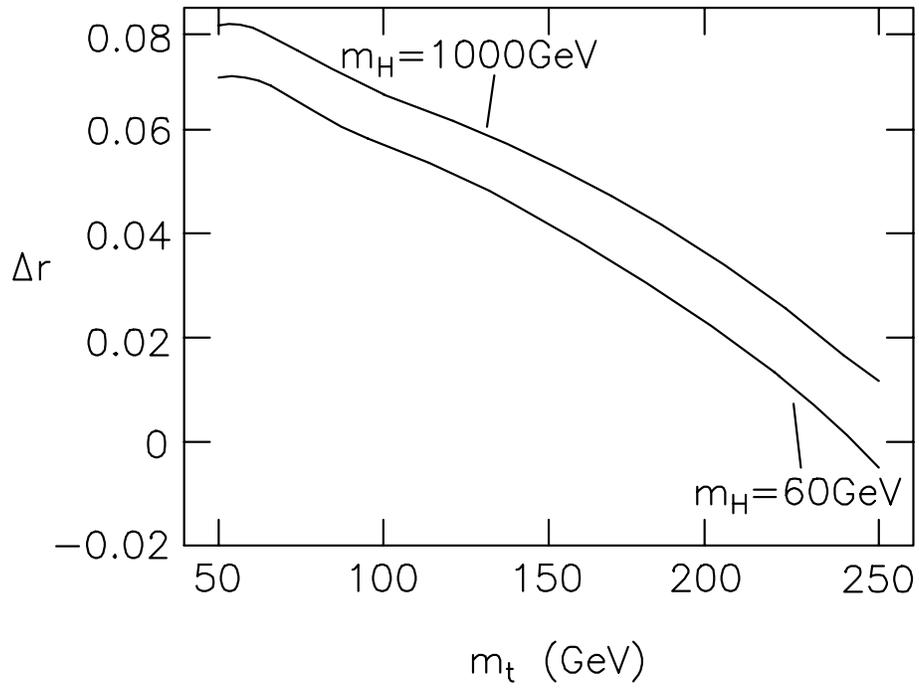

Figure 2: the standard model prediction for $\Delta r$ as a function of the top quark mass

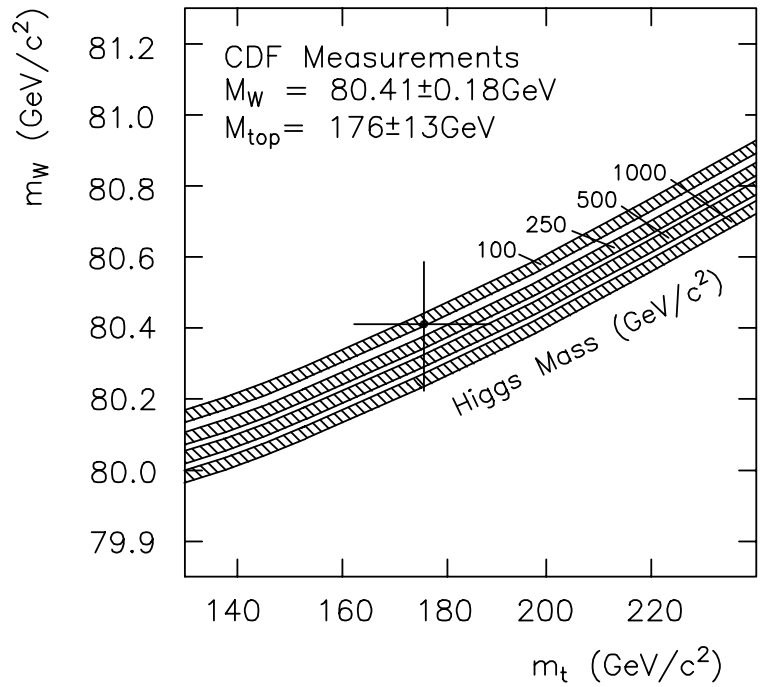

Figure 3: standard model correlations between $m_t$, $m_H$ and $m_W$

The complete dependence of $\Delta r$ on $m_t$ and $m_H$ is shown in fig. 2. With the result of the oneloop calculation one can go back into the defining equation for $\Delta r$ to obtain a curve in the $m_t$–$m_W$ plane (for fixed values of $m_H$). Such curves are shown in fig. 3 for $m_H$=100, 250, 500 and 1000 GeV. They depend much stronger on the precise value of $m_W$ than on $m_t$ and $m_H$, because the latter enter through higher order effects. The dependence on $m_H$ is so weak that it is not possible at the moment to deduct $m_H$ from the measured values of $m_t$ and $m_W$. A reduction in the experimental errors for $m_t$ and $m_W$ by a factor of ten would be helpful, but even this would not allow much more than an estimate of $m_H$. The reason for that is an uncertainty induced on $\Delta r$ by the hadronic uncertainty in $\Delta \alpha$. This uncertainty is reflected in the vertical extensions/shaded regions of the curves in fig. 3. To understand it, one should remember that the running of $\alpha$ is given by the real part of the renormalized vacuum polarization of the photon,

$$\alpha(0) - \alpha(q^2) = \alpha \Pi_{\gamma\gamma}(q^2) \tag{21}$$

The contribution of virtual lepton loops to the vacuum polarization is given by

$$\Pi_{\gamma\gamma}^{leptons}(q^2) = \sum_l \frac{\alpha}{3\pi} \left\{ \frac{5}{3} - \ln(\frac{q^2}{m_l^2}) \right\} \tag{22}$$

A similar formula would hold for light quarks, with $m_l$ replaced by $m_q$, if there would be no confinement of quarks. One cannot calculate $\Pi_{\gamma\gamma}^{quarks}$ but has to cut the vacuum polarization diagram. This allows, via the optical theorem, to relate $\Pi_{\gamma\gamma}^{quarks}$ to the measured values of $\sigma_{total}(e^+e^- \to hadrons)$.

The main contribution to $\sigma_{total}(e^+e^- \to hadrons)$ comes from the low energy region. The low energy resonances are also the main source of error of $\Pi_{\gamma\gamma}^{quarks}(m_Z^2) = -0.0282 \pm 0.0009$ [9]. Using $e^+e^-$ data up to 40 GeV, this number includes effects of all quarks except for the top quark. The contribution from the top quark loop, and in general from heavy particles, to $\Pi_{\gamma\gamma}(m_Z^2)$ is small, of order $O(\frac{\alpha m_Z^2}{m_t^2})$, and can be exactly determined, because the top quark is in some sense a free quark and with the few hundred events from Fermilab the top quark mass is already known more precisely than the masses of u,d and s.

In summary one obtains

$$\Delta \alpha = -\Pi_{\gamma\gamma}(m_Z^2) = 0.0595 \pm 0.0009 \tag{23}$$

The error in $\Delta \alpha$ dominates the theoretical error in the curves fig. 3.

Experimentally, there are two ways to determine $\Delta r$. One can either determine it from the W–mass measurement at Tevatron. This yields $\Delta r_{Tevatron} = 0.040 \pm 0.005$. Or one can try to determine it from the combined LEP1 data. This yields

$$\Delta r = 0.043 \pm 0.0111 \qquad (\Delta r)_{SM} = 0.0396 \pm 0.0035 \qquad (24)$$

The value $(\Delta r)_{SM}$ has been obtained from LEP1 data under the assumption that the Standard Model is correct. Via eq. 15 this corresponds to $m_W = 80.32 \pm 0.06$ GeV.

## 4. Definition of $\sin^2 \theta$ in terms of $A^\mu_{FB}$

The definitions of $\sin^2 \theta$ given above were in fact not so closely related to the LEP1 observables, but rely more on the W mass measurement at Tevatron and to muon decay. A "LEP1 definition" of $\sin^2 \theta$ can be given [10] in terms of the neutral current couplings

$$g_V^f = I_3^f - 2Q_f \sin^2 \theta \qquad g_A^f = I_3^f \qquad (25)$$

whose ratio is measured in forward backward asymmetries. $I_3^f$ and $Q_f$ are the weak isospin and electric charge of fermion flavour f. The forward backward asymmetry in the process $e^+e^- \to Z \to f\bar{f}$ is defined as the difference of events where f goes to the right resp. to the left of the $e^+e^-$ beam, normalized to the total number of f–events. To leading order in the standard model it is given by

$$A^f_{FB} = \frac{3}{4} A_e A_f + O(\frac{m_f}{m_Z}) + O(\frac{\Gamma_Z}{m_Z}) \qquad (26)$$

where

$$A_f = \frac{2\frac{g_V^f}{g_A^f}}{1 + (\frac{g_V^f}{g_A^f})^2} \qquad (27)$$

and is thus a measure of the ratio $\frac{g_V^f}{g_A^f} = 1 - \frac{2Q_f}{I_3^f} \sin^2 \theta$ which arises in the neutral current

$$j^{NC,f}_\mu = \frac{e}{2 \sin \theta \cos \theta} \bar{f}(g_V^f \gamma_\mu + g_A^f \gamma_5 \gamma_\mu) f \qquad (28)$$

The fermion mass term as well as the finite Z width corrections to eq. 26 are explicitly known and can be corrected for. $O(\frac{\Gamma_Z}{m_Z})$ corrections at the Z pole arise, for instance, from $\gamma$–Z interference.

One can use the accurate LEP1 measurements of $A_{FB}$ for muons together with the assumption of lepton universality to define a quantity $s_l^2$ via

$$(\frac{g_V}{g_A})_l = 1 - \frac{2Q_l}{I_3^l} s_l^2 \qquad (29)$$

where $(\frac{g_V}{g_A})_l$ is meant to be the ratio as extracted from the measurement of $A_{FB}^\mu$.

Using polarized electrons the SLC experiment was in fact able to determine $A_e$ separately, and to a precision comparable to the LEP1 result. The two measurements can be combined to give $s_l^2$ with an error of $\pm 0.0003$. With this error $s_l^2$ is a factor of ten more accurate than $s_W^2$ although not as accurately given as $s_0^2$.

In the literature $s_l^2$ is sometimes used to define the $\rho$–parameter, i.e. $s_l^2 = 1 - \frac{m_W^2}{\rho_l m_Z^2}$. This is something like a "generalized" definition of the $\rho$–parameter, but it should be clear that $\rho_l$ is different from the definition of $\rho$ given in eq. 6. It turns out that the leading terms eq. 8 of $\rho$ and of $\rho_l$ agree so that the difference is numerically not so large, but from a principle point of view it is important. Also, one may define a relation between $s_l^2$ and $s_0^2$ by

$$s_l^2 = (1 + \Delta k') s_0^2 \qquad (30)$$

just as a relation between $s_W^2$ and $s_0^2$ was defined in eq. 15. In combination, the three quantities $\Delta k'$, $\Delta \rho$ and $\Delta r$ comprise the complete information hidden in the genuine electroweak oneloop corrections to the process $e^+e^- \to Z \to f\bar{f}$. This statement is true for all $f \neq b$. b quark production involves additional ingredients to be discussed in the next section.

An alternative set of three quantities containing the same information is given by [11]

$$\epsilon_1 = \Delta \rho \qquad (31)$$

$$\epsilon_2 = c^2 \Delta \rho + \frac{s^2}{c^2 - s^2} \Delta r - 2s^2 \Delta k' \qquad (32)$$

$$\epsilon_3 = c^2 \Delta \rho + (c^2 - s^2) \Delta k' \qquad (33)$$

These linear combinations have the advantage that the terms of order $G_F m_t^2$ are concentrated in $\epsilon_1$ and drop out in the combinations $\epsilon_2$ and $\epsilon_3$. Therefore, $\epsilon_2$ and $\epsilon_3$ are dominated by conributions from heavy particles without custodial $SU(2)_R$ breaking. Furthermore, the terms of the form $\ln(m_H)$ appear only in $\epsilon_1$ and $\epsilon_3$, so that $\epsilon_2$ is dominated by terms of the form $\ln(m_t)$. Sometimes in the literature the genuine electroweak oneloop corrections are discussed in terms of an equivalent set of quantities, S,T and U, defined via [12]

$$\epsilon_1 = \alpha T \qquad \epsilon_2 = \frac{-\alpha U}{4s_W^2} \qquad \epsilon_3 = \frac{-\alpha S}{4s_W^2} \qquad (34)$$

In the original paper [12] a truncation to the leading terms in $\frac{1}{m_Z^2}$ was used to define S,T and U which is not really necessary.

All genuine electroweak oneloop corrections to LEP1 observables can be given in terms of $\epsilon_1,\epsilon_2$ and $\epsilon_3$. For example,

$$A_{FB}^l = A_{FB}^l\Big|_{IBA}(1 + 34.72\epsilon_1 - 45.15\epsilon_3) \qquad (35)$$

One can do a combined fit [13] of all the measurements to determine $\epsilon_1,\epsilon_2$ and $\epsilon_3$. I am not going to present the results of this fit here because the particular numbers change from month to month and depend on whether one uses other input than LEP1 or not (like $m_W$ from the Tevatron). Instead I want to finish this section with a qualitative statement on the errors of $\epsilon_1,\epsilon_2$ and $\epsilon_3$. The main error on $\epsilon_1,\epsilon_2$ and $\epsilon_3$ is induced by the errors on $\alpha_s(M_Z^2)$, $\Delta\alpha$, $m_W$ and $m_t$. The first two lead to an ambiguity in the IBA prediction for $A_{FB}^l$, whereas the latter two enter through the uncertainty in $\Delta\rho$. The uncertainty in $\alpha_s(M_Z^2)$ is reflected most strongly in an uncertainty in $\epsilon_3$.

## 5. b Production at LEP1

b production at LEP1 involves a very interesting feature not present in the production of all the other light fermions, namely the presence of vertex diagrams with virtual top quarks in the loops (c.f. fig. 4 and ref. [14] ). It turns out that these corrections affect the integrated width $\Gamma_b = \Gamma(Z \to b\bar{b})$ but not the forward backward asymmetry $A_{FB}^b$. Therefore I shall first take the opportunity to present the formula $\Gamma_f = \Gamma(Z \to f\bar{f})$ for all light flavours f and afterwards discuss the modifications necessary to get $\Gamma_b$. One has

$$\Gamma_f = \Gamma_0(g_V^{f2} + g_A^{f2}(1 - 6\frac{m_f^2}{m_Z^2}))(1 + Q_f^2\frac{3\alpha}{4\pi}) + QCD \qquad (36)$$

where $\Gamma_0 = \frac{1}{12\sqrt{2}\pi}N_c^f G_F m_Z^3$ and "QCD" stand for the QCD corrections, now known including all terms O($\alpha_s^2$). We note that in eq. 36 the two quantities $g_V^f$ and $g_A^f$

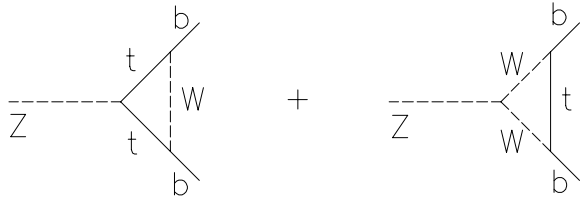

Figure 4: some of the nonuniversal vertex diagrams specific to b–quark production at LEP1

appear separately, and not just their ratio. Beyond the leading order we write

$$g_V^f = \rho_f (I_3^f - 2 Q_f s_f^2) \qquad g_A^f = \rho_f I_3^f \qquad (37)$$

In addition to $s_f^2$, introduced in section 4, a quantity $\rho_f$ appears. For all f except b, one has

$$\rho_f = 1 + \frac{3 G_F m_t^2}{8\sqrt{2}\pi^2} + .... \qquad (38)$$

For f=b one has instead

$$\rho_b = 1 - \frac{G_F m_t^2}{8\sqrt{2}\pi^2} + .... \qquad (39)$$

where the dots stand for nonleading terms. These nonleading terms are known and they are different for $\rho_f$ and $\rho_b$, too. Even twoloop effects (electroweak/QCD interference) are known [15]. In addition, there is a tiny effect from top quark dependent twoloop QCD diagrams specific to b-production [16]. Corrections of the form $\alpha_s G_F m_t^2$ are also known [17]. Note that the diagrams fig. 4 hardly modify the shape of the angular distribution resp. $A_{FB}^b$, but – via eq. 39 – only the integrated width $\Gamma_b$.

Comparing eqs. 38 and 39 we see that these diagrams induce terms of order $G_F m_t^2$ with a tendency to compensate the corresponding contributions from the $\rho$–parameter. This is removed, if one considers the ratio $R_b = \frac{\Gamma_b}{\Gamma_{had}}$ of $\Gamma_b$ to the total hadronic Z width, because in the ratio the "universal", i.e. flavour independent contributions from the $\rho$–parameter drop out. In fact one has

$$R_b = R_b \Big|_{IBA} (1 - 0.06 \epsilon_1 + 0.07 \epsilon_3 + 1.79 \epsilon_b) \qquad (40)$$

with small coefficients of $\epsilon_1$ and $\epsilon_2$. $\epsilon_b$ comprises the effect of fig. 4. At the moment the experimental analysis of b quark production gives values for $R_b$ which are larger than the theoretical prediction (given $m_t$=175 GeV from Fermilab). In terms of top quark mass values they would point more to 120 than to 175 GeV. If this result would persist, this could point to a new physics effect. However, for this analysis an identification of b quarks to per mille accuracy is necessary. Experimentally this seems to be extremely difficult. It is well possible that the explanation of the present discrepancy lies in the misidentification of b and c quarks.

## 6. Conclusions

In this article I have discussed the theoretical implications of the LEP1 electroweak data. I have not discussed possible effects from beyond the Standard Model because my point of view is that one should first analyze the Standard Model predictions very carefully, then compare to the experiments and study other models only in case a deviation from the Standard Model is seen.

At the time of writing the LEP1 measurements are not precise enough to give information on $m_H$ and it is doubtful whether they ever will be.

Certainly, Tevatron and LHC will improve on $m_W$ and $m_t$ and part of the LEP1 precision analysis may become obsolete.It might be just the precise value of $m_Z = 91.1887 \pm 0.0022$, which remains as its most important contribution. Of course, history may evolve differently: There may be something behind the enhanced b production rate. But even in that case, within a few years more precise studies of top quark properties at the Tevatron upgrade could show these effects more prominent.

This is perhaps too pessimistic a picture of an experiment which has fascinated the whole high energy community over many years with its great physics potential and successful data taking. The LEP1 experiment has certainly contributed a lot to our understanding of the particle world and to establish the validity of the electroweak standard model up to energies of order 100 GeV (not to mention the determination of $\alpha_s$ and the number of light neutrino species).

# References

[1] definitions of $\sin^2 \theta$ discussed in original papers are, for example